\documentclass{jfm}
\usepackage{natbib}
\usepackage{epsfig}
\usepackage{amssymb}

\def\begineq{\begin{equation}}
\def\endeq{\end{equation}}
\def\be{\begin{equation}}
\def\ee{\end{equation}}

\title[ Dissipation Scale Fluctuations and Chemical Reaction Rates  in Turbulent Flows.]
{ Dissipation Scale Fluctuations  and Chemical Reaction Rates in Turbulent Flows. }
\author{Victor Yakhot }
\affiliation{`Department of Aerospace and Mechanical Engineering, \\
Boston University, Boston, MA 02215}
\pubyear{2007}
\volume{999}
\pagerange{1--10}
\date{?? and in revised form ??}
\begin{document}
\maketitle




 
\begin{abstract}
\noindent    Small separation between reactants,  not exceeding  $10^{-8}-10^{-7}cm$,  is the necessary condition for various chemical reactions.   It is shown that random advection and stretching by  turbulence  leads  to formation  of   scalar-enriched sheets
 of  {\it strongly fluctuating thickness}   $\eta_{c}$. The molecular-level  mixing  is achieved by  diffusion across these sheets  (interfaces) separating the reactants.  Since diffusion time scale is $\tau_{d}\propto  \eta_{c}^{2}$,  the knowledge of probability density  $Q(\eta_{c},Re)$ is crucial for evaluation of chemical reaction rates.  
 In this paper we derive the probability density $Q(\eta_{c},Re,Sc)$ and predict  a  transition in the reaction rate  behavior from ${\cal R}\propto \sqrt{Re}$ ($Re\leq 10^{4}$) to the high-Re asymptotics ${\cal R}\propto Re^{0}$. The theory leads to an approximate universality of transitional Reynolds number $Re_{tr}\approx 10^{4}$.  It is also shown that if chemical reaction involves short-lived reactants,
 very strong anomalous fluctuations of the length-scale $\eta_{c}$ may lead to non-negligibly small reaction rates.

  \end{abstract}
\section{Introduction}
 \noindent  Efficiency of chemical reactions  and combustion  crucially depends upon number of reactant moles 
mixed on a molecular scale.  Indeed,  for a non negligibly small reaction rate,  the  separation between  reacting species  must not exceed $r\approx 10^{-8}-10^{-7}cm$,  which is the main reason for immense  importance of  mixing process  [1].  Slow diffusion  in laminar media leads to an extremely  poor mixing and, being   the most common mixing accelerator, 
 hydrodynamic turbulence plays vital role 
 in natural and man-made processes  like  heat transfer,  chemical transformations, combustion, meteorology and astrophysics.  
 The mixing  process  in turbulent flows  involves three  main steps: 1.~ entrainment,  creating  pockets of material $B$ in a turbulent flow enriched by  a substance $A$;~2.~advection  and stretching  leading to  formation of  thin   convoluted sheets of the thickness $\eta_{c}$ separating the reactants;   {\it This process is often related to as `mixing by random stirring'.}  3.~ molecular diffusion across  these  `dissipation sheets '  on a time scale $\tau_{d}\approx \eta_{c}^{2}/D$.    If chemical reaction  and  turbulent mixing processes  are fast ~ - ~ molecular diffusion  is the reaction rate - determining  step ( Dimotakis (2005), (1993)).   Recently, investigation of the role  of scalar dissipation and dissipation sheets has become an extremely active field  (see, for example,  excellent reviews by Bilger (2004), Sreenivasan (2004), Peters (2000), Dimotakis (2005) and  recent papers by Celani et al (2005),  Villermaux et al (2003)  and Bush et al (1998). )

\subsection{Kolmogorov ~-~ Batchelor phenomenology.} 
\noindent  Based on a classic Kolomogorov's cascade concept, we  illustrate the main  qualitative features  of the mixing of two liquids (chemical components) $A$ and $B$. (A quantitative dynamic description will be developed below). 
 Consider a turbulent flow of a fluid $A$. The integral scale of turbulence, separating the energy and inertial ranges,  is $L$ and the large-scale Reynolds number is $Re=u_{rms}L/\nu$.   At a time $t=0$, a  blob of the same fluid  enriched with chemical component $B$ is placed in the flow.  For simplicity we assume the linear dimension of a blob $r_{0}=L$. At this point, the area of the interface,  separating $A$ and $B$ substances,  is $O(L^{2})$. 
The life-  (turn-over) time of  this eddy   is $T_{0}\equiv \tau_{eddy}\approx L/u_{rms}\approx L^{\frac{2}{3}}{\cal E}^{\frac{1}{3}}$,  where $u_{rms}$ is the  rms value of turbulent velocity.  The magnitude of the energy flux across scales  is  ${\cal E}=\alpha_{{\cal E}} u_{rms}^{3}/L$ with $\alpha_{{\cal E}}\approx 0.8-0.9$.   In accord with Kolmogorov's phenomenology,     after $\approx T_{0}$ seconds,  by nonlinear interactions,  this  eddy is transformed 
 into another one of the linear dimensions  $L_{x,1}=L/2$ and $L_{y,1}=L_{z,1}\approx \sqrt{2}L$.   Since the gradient in the $x$-direction is  largest,   the characteristic time-scale  of this, ``daughter''  structure 
is:  $T_{1}\approx \frac{L^{\frac{2}{3}}}{{\cal E}^{\frac{1}{3}}}2^{-\frac{2}{3}}$.   Then, after $n\gg 1$ steps
 
 \begin{equation}
 L_{x,n}= 2^{-n}L;~~~T_{n}\approx  2^{-\frac{2n}{3}}\frac{L^{\frac{2}{3}}}{{\cal E}^{\frac{1}{3}}};~~~\tau_{v,n}\approx 2^{-2n}\frac{L^{2}}{\nu}
 \end{equation}

\noindent  where  the viscous time is denoted as $\tau_{v}$. We see that the viscous time of  a  structure  strongly decreases with the number of `cascade steps'.  The time needed to form the smallest Kolmogorov  eddies on the scale $\eta_{K}\approx L Re^{-\frac{3}{4}}$ is thus:

\begin{equation}
T_{K}=T_{0}+T_{1}+\cdot\cdot\cdot T_{n}\approx 2.7 T_{0}\approx \frac{L^{\frac{2}{3}}}{{\cal E}^{\frac{1}{3}}}
\end{equation} 

\noindent  where $n\approx \frac{3}{4}\frac{\ln Re}{\ln 2}\approx \ln Re\gg1$.  It  is important that the Reynolds number $Re=u_{rms}L/\nu$    used in the above relations  is to be distinguished from $Re_{U}=UL/\nu$ where $U$ is the mean velocity in the flow.  Typically,  numerically,   $Re<Re_{U}$.  
The fact that Kolmogorov's  scale  $\eta_{K}$ and dissipation rate  ${\cal E}$  are formed on the time-scale of a single large-scale turn-over time  ($\tau_{eddy}$)  has been tested in various  numerical experiments.

\noindent  If  the Schmidt number $Sc=\nu/D\approx 1$,  in accord with the cascade picture,  the Kolmogorov scale is the smallest legth-scale created by turbulence and  at times $t>T_{K}$,  the mixing proceeds by molecular diffusion. 

\noindent If however,  $Sc\gg1$ and  the scalar diffusion is extremely  inefficient,   then,  after  formation of the Kolmogorov scale $\eta_{K}$,  the stretching process by the large-scale ($r<<\eta_{K}$) velocity field  leads to generation  of the ever thinner scalar-enriched  sheets until  $\eta_{c}(t)\approx \eta_{B}\approx \eta_{K}/\sqrt{Sc}\ll \eta_{K}$.     It is only after that,  the scalar diffusion takes over the mixing process. 
The  mean width of these sheets  is:  $\eta_{B}\approx \eta_{K}/\sqrt{Sc}$, where  $\eta_{B}$ is called the  Batchelor  (1959) scale.  At this stage,   the substances $A$ and $B$,  separated by the distance $\eta_{B}$,  can mix  only  by molecular diffusion  on a time scale $\tau_{d}\approx \eta_{B}^{2}/D\approx \eta_{K}^{2}/\nu\approx \frac{L^{2}}{\nu} Re^{-\frac{3}{2}}$
.  
 
\noindent To get a feel for  ``real life charm'',   we consider a simple numerical example.  In gases where $Sc\approx 1$, the length-scales $\eta_{K}\approx \eta_{B}$.  In liquids,  the situation is different and $Sc\approx 600-3000$ (Dimotakis (2005), (1993)) which means that $\eta_{K}/\eta_{B}\approx 25~-~50$.   Thus, while the scales $\eta_{B}$ and $\eta_{K}$ can be very different,  the corresponding  diffusion and viscous times scales  $\tau_{d}\approx \eta_{B}^{2}/D\approx \eta_{K}^{2}/\nu\approx \tau_{v}$ are of the same order. 
Moreover,  it has been shown in various numerical and experimental studies  (see for example Monin and Yaglom (1975)) that  the relevant viscous scale  $\eta_{\nu}$ is  numerically larger than  Kolmogorov's  one:  $\eta_{\nu}=\alpha_{\nu}  \eta_{K}$ with $\alpha_{\nu}\approx 10-20$.   Thus,   the diffusion time is,  in fact,  $\tau_{d}=\eta_{\nu}^{2}/\nu \approx \tau_{v}$. 
(The origin of the large factor $\alpha_{\nu}$ will be discussed below.) 

\subsection{Low and large Reynolds number behavior  of  reaction rates.}

\noindent  If $Sc>>1$, the  process consists of two steps: 1. Formation of  structures (sheets)  on the viscous scale $\eta_{\nu}\approx \alpha_{\nu} LRe^{-\frac{3}{4}}$;~2.~ Further stretching of the scalar fieled toward $\eta_{c}\approx \alpha_{\nu}\eta_{K}/\sqrt{Sc}$. It can be shown readily (see Monin and Yaglom (1975)) that,  after  initial  formation of the dissipation scale $\eta_{\nu}$,  the distance between two particles across the sheets,    stretched by the large-scale velocity field  decreases with time as  $r(t)=\eta_{\nu}\exp(-\gamma t)$. Thus, the scalar dissipation scale is formed on a time - scale $\tau_{c}$ given by the relation: 

$$\eta_{\nu}e^{-\gamma \tau_{c}}\approx \eta_{\nu}/\sqrt{Sc}$$

\noindent where $\gamma=a_{\gamma}\sqrt{\frac{{\cal E}}{\nu}}$ with $\frac{1}{\sqrt{3}}>a_{\gamma}>\frac{1}{2\sqrt{3}}$. 
This gives:

$$\tau_{c}= \frac{L}{2a_{\gamma}  u_{rms}}Re^{-\frac{1}{2}} \ln Sc$$

\noindent The diffusion time across  scalar dissipation sheets is:

$$\tau_{d}\approx \frac{\alpha_{\nu}^{2} L^{2} Re^{-\frac{3}{2}}}{\nu}\approx \alpha_{\nu}^{2}\frac{L}{u_{rms}}Re^{-\frac{1}{2}}$$

\noindent Comparing the above relations we come to a non-trivial conclusion:   although  the characteristics times  scale with the Reynolds number as $\tau_{c}\propto \tau_{d}\propto \sqrt{Re}$,  the   stretching process leading to  both sheet thinning and  increase of the scalar gradient   competes with the  simultaneous concentration-gradient-decreasing  diffusion  across   the developing interfaces. 
 However,  due to the large magnitude of  numerical factor $\alpha_{\nu}\approx 10-20$,  the process of the sheet-thinning by the large-scale stretching is numerically much faster than  diffusion across the sheets. This means that during the stretching stage,  the scalar diffusion across the interfaces can be neglected.    {\it The  factor $\alpha_{\nu}\approx 10-20$ will be calculated below and  it will become clear that its large magnitude   is a consequence of  complex small-scale dynamics of  intermittent turbulence.}

\noindent Thus, the ratio  of the mixing (inviscid) and diffusion times can be estimated as:

$$\frac{\tau_{eddy}}{\tau_{d}}\approx  \frac{\sqrt{Re}}{\alpha_{\nu}^{2}}$$

\noindent and  $\tau_{eddy}/\tau_{d}\approx 1$ for $Re\approx \alpha_{\nu}^{2}\approx 10^{4}$.

\noindent   In the flows with $Re<10^{4}$, molecular diffusion across the dissipation sheets is the longest,  rate-determining,  process.  In the wall flows,  this Reynolds number corresponds to    $Re_{U}\leq 2-3 \times 10^{5}$.  
 It is only when  $Re_{U}\gg 10^{5}-10^{6}$,  the  rapid  diffusion through extremely thin interfaces,  is dynamically irrelevant for the mixing process and  inviscid mixing time $\tau_{eddy}$  is the rate-determining step.     Based on the above considerations, one can expect  a transition from  the diffusion-  to   advection  - dominated  mixing  at 

\begin{equation}
Re_{tr}\approx \alpha_{\nu}^{4}
\end{equation}

\noindent If  mixing is the reaction rate ( ${\cal R}$ ) determining  process, we expect 

\begin{eqnarray}
{\cal R}\propto \frac{\nu}{\eta_{K}^{2}}\approx  \frac{\nu}{L^{2} }Re^{\frac{3}{2}} = \tau_{eddy}\sqrt{Re}
~~~~~Re< Re_{tr}\nonumber \\
{\cal R} \propto \frac{u_{rms}}{L}\approx  \frac{\nu}{L^{2}}Re^{1}= \tau_{eddy}Re^{0}~~~~~Re>Re_{tr}
\end{eqnarray}

\noindent In accord with (1.4), the transitional Reynolds number depends only upon coefficient $\alpha_{\nu}$,  characterizing   small-scale properties of turbulence. Thus, we can conclude that, {\it since the small-scale property of turbulence $\alpha_{\nu}\approx 10-20$ is a more or less universal number, independent upon type  of the flow, the derived $Re_{tr}\approx 10^{4}$  must be approximately universal. }  A mixing transition leading to the Reynolds - number - independent reaction rate    at  approximately universal Reynolds number $Re_{tr}\approx 10^{4}$,  has been observed in experiments by Dimotakis (2005), (1993).


 \section{Statistical Description of Disipation Structures.  Evaluation of  $\alpha_{\nu}$. }
 
\noindent  The Kolmogorov theory (K41),  treating the u.v. cut-off $\eta_{K}\approx L Re^{-\frac{3}{4}}=const$,  completely   disregarded  the  non-trivial  dynamics of the  dissipation range fluctuations.  It became clear recently that the dissipation scale is not a constant number  but a random field  defined  as
 :

\begin{equation}
\eta\approx \frac{\nu}{\delta_{\eta}u}\equiv \frac{\nu}{|(u(x+\eta)-u(x))|}
\end{equation}

\noindent  where $\delta_{y} u_{i}=u_{i}(x+y)-u_{i}(x)$.  In this form the relation for $\eta$ was derived (Yakhot (2003, (2006)),  Yakhot and Sreenivasan (2004), (2005)) from the dissipation anomaly, first introduced by Polyakov (1995)  for the case of Burgers turbulence  and later generalized to the Navier-Stokes turbulence by Duchon and Robert (2000), Eyink (2003). Even earlier, the fluctuating dissipation scale was used  by Paladin and Vulpiani (1987) in the context of multifractal theory The  local value of the Reynolds number $Re\approx u_{\eta}\eta/\nu\approx 1$, with $\delta_{\eta}u\equiv u_{\eta}$ as a typical speed of an eddy of linear dimension $\eta$,   was mentioned  in Landau and Lifshitz as a criterion for  the onset of viscous dissipation,  as early as 1959.

\noindent The physical meaning of the dissipation scale $\eta$ is understood as follows. As $r\rightarrow 0$, the velocity field is analytic, so that $\delta_{r}u\approx \partial_{x}u(x)r$. Thus, in this limit
$S_{n}=\overline{(\delta_{r}u)^{n}}\propto r^{n}$. On the other hand, in the inertial range $S_{3}\propto r$ 
which, contradicting small-scale analyticity of velocity field,  cannot be valid in the limit $r\rightarrow 0$. 
Moreover,  in isotropic and homogeneous turbulence, the dissipation rate ${\cal E}\propto \nu\overline{(\partial_{x}u_{x})^{2}}=-\nu\lim_{r\rightarrow 0}\frac{\partial^{2}}{\partial r^{2}}S_{2}(r)\propto \nu \lim_{r\rightarrow 0} r^{-\frac{4}{3}}=O(1)$ is regularized by viscosity $\nu\rightarrow 0$. This leads  to the  definition of the dissipation scale $\eta_{K}$ :  ${\cal E}^{\frac{1}{3}}\propto \nu \eta_{K}^{-\frac{4}{3}}$ and, since ${\cal E}\approx u_{rms}^{3}/L$,  to Kolmogorov's estimate for $\eta_{K}\propto Re^{-\frac{3}{4}}$.  

\noindent Based on the above considerations,  defined by (2.1),  $\eta$ is a scale separating analytic ($r\ll \eta$)  and singular ($r\gg \eta$)  contributions to turbulent  velocity field.  The relation (2.1) is an order of magnitude estimate of the "dissipation scale" and,  in general,  $\eta\approx \frac{a \nu}{\delta_{\eta} u}$ where $a$ is a velocity- field -independent  factor which was investigated in numerical simulations by J. Schumacher  et. al. (2007) .  Interested in qualitative aspects of the mixing process, we, for now,  neglect this coefficient and use expression  (2.1).

\noindent  According to the theory (Yakhot (2003)),  there exist an infinite number of ``dissipation scales '' $\eta_{n}$ separating smooth $S_{n}\propto r^{n}$    ($r\ll \eta_{n}$)   and singular $S_{n}\propto r^{\xi_{n}}$   ($r\gg \eta_{n}$)  intervals of the moments $S_{n}=\overline{(\delta_{r} u)^{n}}$.  This fact has been decisively demonstrated in numerical experiments of Schumacher et al (2007).  According to the analytic theory (Yakhot (2003)), in general 

$$L Re^{-\beta}\leq \eta_{n}\leq L Re^{-\frac{1}{2}}$$

\noindent where $\beta\approx 1$ and the Kolmogorov scale  $\eta_{K}\approx \eta_{2}$ is only one of the possible dissipation scales.  This property of turbulence will be important in what follows.

\noindent Let us consider an equation for concentration  $c$ of a passive scalar advected by velocity ${\bf u}$:

\begin{equation}
\frac{\partial c}{\partial t}+{\bf u \cdot \nabla}c=D\nabla^{2}c;  \hspace{0.5in} {\bf \nabla\cdot u}=0
\end{equation}

\noindent   
We assume that the velocity field ${\bf u}$ is   governed by the Navier-Stokes equations and 
${\bf u=U+u'}$ with ${\bf U}$ and ${\bf u'}$ corresponding to quasy-regular (sometimes time-dependent)  and chaotic (turbulent) contributions, respectively.   The fluctuations of the scalar dissipation field  $N=D(\nabla c)^{2}$,  governed by (2.2),   have been investigated in  great detail  for the case of the $\delta$-correlated in time,  large-scale,  velocity field (Kraichnan-Batchelor problem),  where  the stretched exponential  tail of the distribution function  $P(N)$  was derived in the range $N>>\overline{N}$ ( Chertkov et. al (1998), Gamba et al (1999)).    In this paper we are interested in statistical properties  of  ``molecular diffusion  sheets'',  which are  not directly related to the tails of the scalar dissipation rate distribution. 

\noindent The scalar field is analytic,  so that for $r\rightarrow 0$, $c(x+r)-c(x)\approx \frac{\partial c(x)}{\partial x}r$.  In addition,  in the scalar ``inertial range''   $r\gg \eta_{K}$ (Monin and Yaglom (1975):
 
\begin{equation}
S_{3}^{u,c}=\overline{(u(x+r)-u(x))(c(x+r)-c(x))^{2}} =-\frac{4}{3}\overline{N}r
\end{equation}

\noindent  where $N=D\overline{(\frac{\partial c}{\partial x_{i}})^{2}}=O(1)$.  In the case $Sc>>1$, there exist an additional scalar ``rough''  range $\eta_{B} \leq r\leq \eta_{K}$, where $S_{3}(r)\propto \ln r$. It is only at the scales $r\ll \eta_{B}$, the scalar field is smooth. 
By definition,  the length scale $r\approx \eta_{c}$  is the scale separating analytic and singular contributions to the scalar field $c({\bf r},t)$.  As follows from  (2.3),  in the inertial range, the scalar field  $c({\bf x})$ is  not differentiable and in the limit $D\rightarrow 0$,  one has to be careful with evaluation of  spatial derivatives of the scalar field. 

\noindent   From the equaion (2.2) we have 

\begin{equation}
\frac{\partial c^{2}}{\partial t}+{\bf u \cdot \nabla}c^{2}=2Dc\nabla^{2}c
\end{equation}

\noindent and introducing the ``point - splitting''  $c(\pm)=c(x\pm y)$ and ${\bf u}(\pm)={\bf u}(x\pm y)$, derive:
\begin{equation}
\frac{\partial (c(+)c(-)}{\partial t}+(\nabla_{+}{\bf u}(+)+\nabla_{-}{\bf u}(-))c(+)c(-)=D(\nabla_{+}^{2}+\nabla_{-}^{2})c(+)c(-)
\end{equation}

\noindent  {\it Equation (2.5) involves derivatives of  singular ( in the inertial range)  functions.  Thus,  in the limit $y\rightarrow 0$, the exact equation  (2.4) for the scalar variance can appear  from (2.5)  only if singular and regular contributions balance separately.}   Indeed, 
in the limit $y\rightarrow \eta_{c}\rightarrow 0$,  taking into account that $\frac{\partial}{\partial y}=\frac{\partial}{\partial x_{+}}=-\frac{\partial}{\partial x_{-}}$ and repeating all steps presented in detail in (Yakhot (2006)),  we derive:

\noindent 

\begin{equation}
\frac{\partial}{\partial y_{i}}(\delta_{y}u_{i}(\delta_{y}c)^{2})+2\nabla_{+}{\bf u}(+)c^{2}(-)+2\nabla_{-}{\bf u}(-)c^{2}(+)=-4D\delta_{y}c\frac{\partial^{2}}{\partial y^{2}}\delta_{y}c
\end{equation}

\noindent The   relation (2.6) is  exact locally  in space and time.  Since in incompressible isotropic turbulence the velocity- scalar correlation function  $\overline{u_{i}(x)c^{2}(x')}=0$ (Monin and Yaglom (1975)),  it is clear that averaging over a "ball" of radius $\eta_{c}$ (Duchon (2000), Eyink (2003),   the second and third terms in the left  side of (2.6) disappear 
giving the locally valid estimate for the scalar dissipation scale,  independent upon specific model of turbulence:

\begin{equation}
\eta_{c}\approx \frac{D}{\delta_{\eta_{c}}u}
\end{equation}

\noindent As follows from this relation,  {\it the random variable}  $\eta_{c}$  depends upon local values of velocity fluctuations.  
In the most interesting  and important case $Sc=\nu/D\gg1$,  on the scale $\eta_{c}\ll \eta$ the velocity field is analytic, giving:

\begin{equation}
\eta_{c}^{2}\approx \frac{D}{\frac{\partial u(x)}{\partial x}}\approx \frac{D \eta}{\delta_{\eta} u}\approx \frac{D\eta^{2}}{\nu}=\eta^{2}/Sc
\end{equation}

\noindent Thus,  the probability of the scalar dissipation scale $Q(\eta_{c})$ is evaluated readily from the PDF $Q(\eta)$ calculated in Yakhot (2006). 
This result leads to some important consequences: If  

\begin{equation}
\tau_{d}\approx \eta^{2}_{\nu}/\nu=\alpha_{\nu}^{2}\eta_{K}^{2}/\nu\approx \alpha_{\nu}^{2}\tau_{eddy}/\sqrt{Re}
\end{equation}

\noindent      to evaluate the reaction rate and the proportionality coefficient $\alpha_{\nu}$, we need the probability  density function $Q(\eta_{c})=\sqrt{Sc}Q(\eta  \sqrt{Sc})$.

\section{Probability densities.}

 \noindent  In what follows we set $L=1$, so that $\frac{r}{L}\equiv r<1$ and if  the moments of velocity increments $S_{n,0}=\overline{(\delta_{r}u)^{n}}=A(n)r^{\xi_{n}}$,  
then  the 
probability density function can be found from the  Mellin transform:

\begin{equation}
P(\delta_{r}u,r)=\frac{1}{\delta_{r} u}\int_{-i\infty}^{i\infty} A(n)r^{\xi(n)}(\delta_{r}u)^{-n}dn
\end{equation}

\noindent   where 
we set the integral scale $L$ 
and the dissipation  rate ${\cal E}$ equal to unity. Indeed, multiplying (3.1) by $(\delta_{r}u)^{k}$ and evaluating a simple integral, gives  $S_{k,0}=A(k)r^{\xi_{k}}$.  Under different name this transformation has been used  in  Tcheou et.al  (1998) in the context of  multifractal theory of turbulence . 
With the Gaussian large-scale boundary condition for the probability density at $r=1$,  the amplitudes $A(n)=(2n-1)!!$  and,   for the values of $n<1/b$, we can use the Taylor expansion of the function $\xi_{n}$  giving  $\xi_{n}\approx( an-bn^{2})$.  The detailed theory for this case has been recently developed in Yakhot (2006)   with the result  ($\delta_{\eta}u\equiv u$):

\begin{equation}
P(u,r)=\frac{2}{\pi u\sqrt{4b| \ln r|}}\int_{-\infty}^{\infty}e^{-x^{2}}
exp[-\frac{ (\ln \frac{u}{r^{a}\sqrt{2}x})^{2}}{4b|\ln r|}]dx
\end{equation}

\noindent The PDF  $P(u/r^{a},r)$ is plotted on Fig. 1  for a few values of the displacement $r$.   Here
$a\approx 0.38$ and $b=0.017$. On Fig. 1b we see the broad tails which, as  a sign of  strong intermittency, 
cannot be collapsed on a single curve.

\begin{figure}[h]
  {\includegraphics[height=4cm]{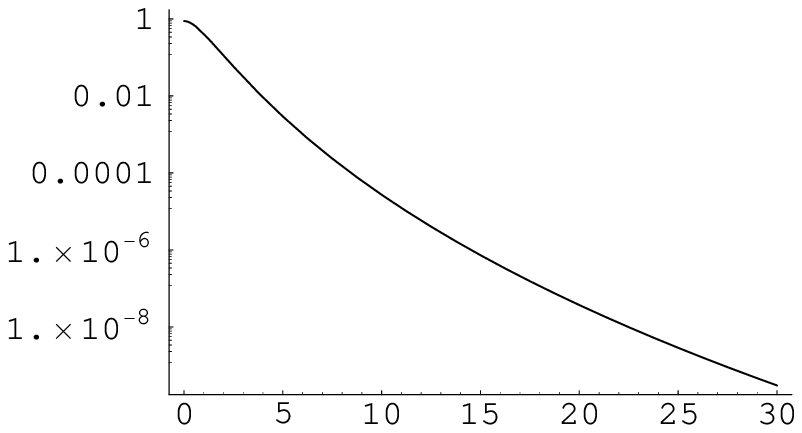}}
  {\includegraphics[height=4cm]{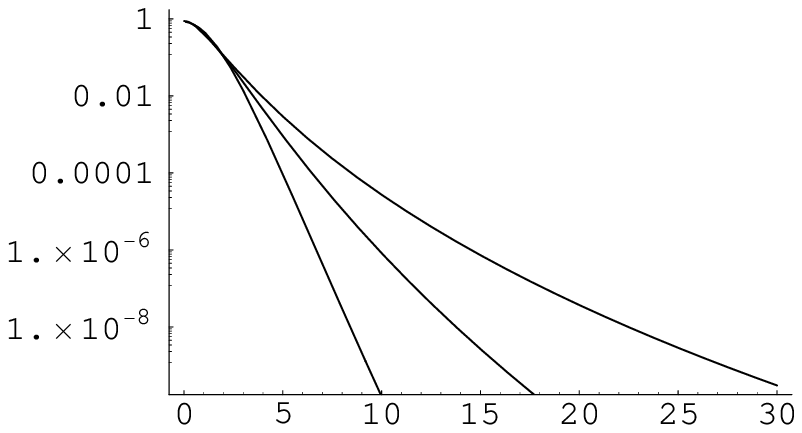}}
  \caption{Inertial range: $r\gg \eta_{0}$. Probability density $P(\frac{\delta_{r}u}{r^{a}},r)$ vs $\delta_{r}u/r^{a}$. a.~r=0.01; b~r=0.01;~0.1;~0.5}
\end{figure}

\noindent According to its definition,  the dissipation scale is a linear dimension of a structure defined by the  local value of the Reynolds number $Re_{\eta}=\eta \delta_{\eta}u/\nu=O(1)$.  Experimentally,  probability density $Q(\eta,Re,Re_{\eta})$ is found by fixing the displacement $r=\eta$ and counting the  events with $\eta\delta_{\eta}u/\nu=Re_{\eta}=a$. This algorithm has been used by Schumacher (2007) in his numerical investigations (see Fig. 3).  From the formula  (3.1) we have: 
 
\begin{equation}
P(\delta_{\eta} u)\equiv P(u_{\eta})=\frac{1}{u_{\eta}}\int_{-i\infty}^{i\infty} A(n)\nu^{\xi(n)}u_{\eta}^{-\xi(n)-n}dn
\end{equation}
\noindent  Fixing $L={\cal E}=u_{rms}=1$ gives  the large-scale Reynolds number 
$\nu=1/Re$   and taking into account that by virtue of (2.1) $ u_{\eta}Re/u_{rms}=\frac{L\delta_{\eta}u}{\nu}\approx L/\eta$ gives  for the probability density $Q(\eta)$ (in what follows we denote $\frac{\eta}{L}\equiv \eta$):

\begin{equation}
Q(\eta,Re)=\frac{1}{\eta}\int_{-\infty}^{\infty}e^{-x^{2}}dx\int_{-\infty}^{\infty}dn
 e^{in \ln(\eta^{a+1}\sqrt{2}xRe)-bn^{2}\ln\eta}
\end{equation}
\noindent and:

\begin{equation}
Q(\eta,Re)=\frac{1}{\eta\sqrt{4b\ln\eta}}\int_{-\infty}^{\infty}
e^{-x^{2}}dx e^{-\frac{\ln^{2}(\eta^{a+1}\sqrt{2}xRe)}{4b\ln \eta}}=
\frac{1}{\eta \sqrt{4b|\ln\eta|}}\int_{-\infty}^{\infty}
e^{-x^{2}}dx e^{-\frac{\ln^{2}((\frac{\eta}{\eta_{0}})^{a+1}\sqrt{2}x)}{4b|\ln \eta|}}
\end{equation}

\begin{figure}
 {\includegraphics[height=6cm]{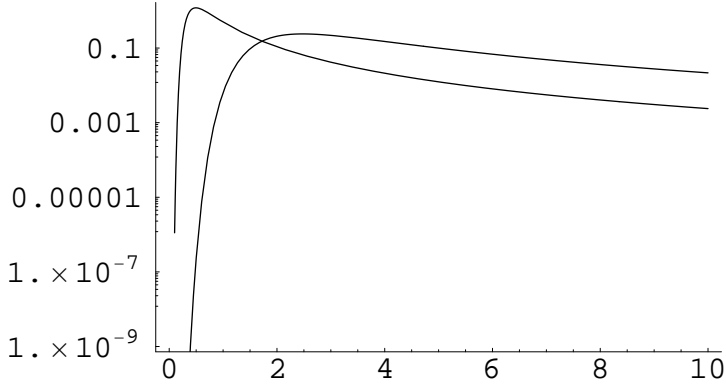}}
  \caption{ Logarithm of probability densities   of  the scalar dissipation scale  $Q(\frac{\eta_{c}}{\eta_{0}},Re,Sc)$ ;  $Re=10^{4}$, $Sc=1$ (right curve), $Sc=25$ (left curve). 
  $\eta_{0}=L Re^{-\frac{1}{1+a}}$ with $a=0.38$. }
\end{figure}


\noindent where $\eta_{0}=LRe^{-\frac{1}{1+a}}$. As expected, the probability density of  dissipation scales is expressed in terms of the ratio $\eta'=\eta/\eta_{0}$ and the width of the distribution is the weak function of the Reynolds number.  The PDFs $Q(\eta/\eta_{0},Re)$ for  $Re=10^{4}$ is shown on Fig. 2 for two values of the Scmidt number $Sc=1$ and $c=25$. 

\section{ Mean dissipation scale and diffusion time.} 

\noindent Now we can evaluate  the moments of the dissipation scale:
\begin{equation}
\overline{e^{n}}\equiv \overline{(\frac{\eta}{\eta_{0}})^{n}}=\int_{0}^{\infty}( \frac{\eta}{\eta_{0}})^{n}Q(\eta, \eta_{0})d\eta
\end{equation}
\noindent and mean diffusion time:
\begin{equation}
\overline{\tau_{d}}=\overline{\eta^{2}}/\nu=\int_{0}^{\infty}\eta^{2}Q(\eta,\eta_{0},Re)\ d \eta /\nu
\end{equation}

\noindent The numerical results slightly vary  with  position of the maximum of  PDF $Q(\eta/\eta_{0})$,  which depends upon the magnitude of parameter $a$ in the expression  $a\eta u_{\eta}/\nu=1$.   If we choose $a$ so  that, in accord with numerical simulation of Schumacher (2007),  the maximum is  set at $\eta_{max}/\eta_{0}\approx 2$, then numerical integration (4.1) yields $\overline{\eta}=\alpha_{\nu}\eta_{0}\approx 7\eta_{0}$ and $\overline{\eta^{2}}\approx 120\eta_{0}$.  If however,  $\eta_{max}\approx 5\eta_{0}$,  we derive $\overline{\eta }=\alpha_{\nu}\eta_{0}\approx 13.5\eta_{0}$, which is close to the outcome of Dimotakis's (2005)  physical  and Gotoh-Nakano's   (2003) numerical experiments.  In this case, $\overline{\eta^{2}}\approx 500\eta_{0}$.  In general, based on (4.1),  (3.5), $\overline{\eta^{2}}\gg \alpha_{\nu}^{2}\overline{\eta}^{2}$.
This result leads to important conclusion: {\it Due to strong intermittency of the dissipation scales,   the $O(\overline{\eta^{2}}/\nu)$ scalar diffusion time is much  longer than  $\tau_{d}\approx \alpha_{\nu}^{2}\eta_{K}^{2}$,  calculated on the basis of Kolmogorov's phenomenology.}  Therefore,  molecular  diffusion,   as a reaction rate determining process,  is even more restrictive  than previously thought.  We  can also conclude that these fluctuations are responsible for the ``large''  magnitude  of a constant $\alpha_{\nu}\approx 10-20$.

\subsection{Mixing  reactants having finite life-time .}

\noindent One can define  dimensionless  Damk\"oller number $Da=(\tau_{c}+\tau_{d})/\tau_{reaction}$,  which is the ratio of hydrodynamic mixing  time scale to $\tau_{reaction}$,  characterizing the reaction rate between  perfectly mixed reactants $A$ and $B$.  In case of fast reactions,  we are interested in here,  $Da>>1$.  Below, we will show that in some cases,  to describe chemical reactions,  Kolmogorov's  cascade picture,  though elegant and illuminating,   is not sufficient. 

\noindent In what follows  we consider a simple example of a model photo-chemical reaction $A^{*}+B=AB+h\nu$ where $A^{*}$  is a component $A$ in an initially prepared electronically excited  state characterized by  a life-time $\tau_{e}$. By definition, the finite life-time $\tau_{e}$ implies time-dependence of concentration of excited states  $c_{A^{*}}=c(t=0)\exp(-\frac{t}{\tau_{e}} )$.  As above,  we are interested in a diffusion-dominated limit  $Re \leq 10^{4}$. 
A chemical reaction is  possible only if  diffusion  time $\tau_{d} \approx  \alpha_{\nu}^{2}\eta_{B}^{2}/D \leq \tau_{e}$,  $\alpha_{\nu}\eta_{B}\leq \sqrt{\tau_{e}D}$ and  $\eta_{\nu}\leq \sqrt{\tau_{e}\nu}$.  This gives,  in addition to the Damk\"ohller mumber, 
a dimensionless  reaction criterion:

\begin{equation}
Y=\alpha_{\nu}\sqrt{\frac{\tau_{eddy}}{\tau_{e}}}Re^{-\frac{1}{4}}\leq 1
\end{equation}

\noindent When the life-time $\tau_{e}$ is very small,   according to the relation (4.3),  which is the outcome of Kolmogorov's phenomenology,  the photo-chemical reaction is impossible.   Now we will show that due to strong  fluctuations 
of the scalar dissipation scale $\eta_{c}$, this conclusion must be modified. 

\noindent      Let us  consider the advection-diffusion equation for concentration of a passive scalar $c\equiv c_{A^{*}}$  undergoing a chemical reaction with another  one of concentration $c_{B}$. If  the maximum separation,  for which reaction is still possible (reaction radius)  is $\Delta$, we can define the probability to find a molecule  $B$ within a  sphere of radius  $\Delta$ surrounding the reactant $A^{*}$ as ${\cal P}(r<\Delta)$ and write the balance equation for the reactant $A$ as: 
\begin{equation}
\frac{\partial c}{\partial t}+{\bf u \cdot \nabla}c=D\nabla^{2}c-c {\cal P}(r\leq \Delta)w_{AB}-\frac{c}{\tau_{e}};  \hspace{0.5in} {\bf \nabla\cdot u}=0
\end{equation}

\noindent where $w_{AB}$ is the reaction rate of perfectly mixed reactants separated by the distance $r\leq \Delta$. The $O(1/\tau_{e})$ term in (4.4) accounts for the finite life -time of one of one  the reactants  $A^{*}$.  {\it It is clear that ${\cal P}(r\leq \Delta)$ depends upon concentration $c_{B}$. }  As was mentioned above,  the reaction radius  $\Delta$, depending on the overlap of molecular orbitals,  is very small and,  during the mixing stage,  when  $r(t)>>\Delta$,  the probability ${\cal P}=0$ and the $O(w_{AB})$ chemical  contribution to the balance equation (4.4) can be neglected. The relation (4.4) illustrates importance of the molecular-level mixing process in chemical kinetics.

\noindent By substitution  $c_{A}\equiv c\rightarrow ce^{-\frac{t}{\tau_{e}}}$,  the remaining equation  is transformed into (2.2).  It is clear  that the reaction rate is not negligibly small only if the mixing time $\tau_{d}<\tau_{e}$.   

We illustrate the qualitative features of the process on a numerical example. 
If $\overline{\eta}/\eta_{0}\approx 13$, (Gotoh/Nakano (2003)),  then, based on the PDFs computed above,  $\overline{\tau_{d}}/\tau_{0}\approx 500$. 
For  $\tau_{e}/\overline{\tau_{d}}\approx g\ll 1$, the  naive (mean-field ) reaction yield, proportional to the concentration  $c\propto \exp(-\overline{\tau_{d}}/\tau_{e})$ , is negligibly small. 
However,  defining $\tau_{e}\approx \eta_{e}^{2}/\nu$,  gives  $\frac{\eta_{e}}{\eta_{0}}\approx \sqrt{500 g}$. 
This result means that a finite  a fraction 

\begin{equation}
F=\int_{0}^{\eta_{e}}Q(\eta,\eta_{0},Re)d\eta
\end{equation}

\noindent  of dissipation sheets with $\eta\leq \eta_{e}\approx   \sqrt{500 g}\eta_{0}$ does contribute to the nonzero reaction rate.  Since $\tau_{d}\approx \tau_{v}$,  the integral is evaluated using the probability density of velocity dissipation scales.

\noindent Taking, for example,  $g\approx 0.02$, we see that only the sheets of the thickness $\eta<\sqrt{10}\eta_{0}\approx 3\eta_{0}$ contribute to this  reaction.  The fraction of the dissipation structures satisfying this condition is $F\approx 0.3$.  If $g\approx 0.01$,  we find $F\approx 5\times 10^{-4}$.  We can see that, due to strong fluctuations of the dissipation scale,  the reaction is not negligible even when $\tau_{e}\ll \overline{\tau_{d}}$.  

\section{Conclusions.} 

 \noindent  To conclude:  the  scalar and velocity dissipation scales $\eta$ and $\eta_{c}$ in turbulent flows are not constant numbers but  describe  random fields with 
 $\eta_{c}\approx D/\frac{\partial u}{\partial x}$ and $\eta\approx \nu/\delta_{\eta}u$, respectively. In an important case $Sc\gg1$, these scales are  related as  $\eta_{c}\approx \eta/\sqrt{Sc}$. Based on the Mellin transform  and Taylor expansion of the scaling exponents of  velocity structure functions, the probability density of the scalar dissipation scale has been derived.  Two main   results of this paper are:  1.~ Due to strong small-scale intermittency, the calculated mean thickness of a dissipation sheet is $\alpha_{\nu}\eta_{K}$ where $\alpha_{\nu}\approx 10-13$.  Extremely strong intermittency  leads to a long  scalar- transport  time $\overline{\tau_{d}}$ across the sheets and, in the flows with $Re\leq \alpha_{\nu}^{4}\approx 10^{4}$,  to   diffusion as a reaction rate- determining step . Therefore, the reaction rate is: ${\cal R}\propto \sqrt{Re}$ for $Re<10^{4}$, and ${\cal R}\propto Re^{0}$ in the interval $Re>10^{4}$. ~ 2.~ Even when the life - time of the reactants is very short,  due to the dissipation scale fluctuations,  the reaction can proceed via diffusion across  thinnest  dissipation sheets $\eta_{c}<<\overline{\eta_{c}}$. In this case, the fluctuations  lead  to the non-negligibly  small reaction rates.   
This result may be  of importance for reactions involving short-lived radicals, excimers and other cases. We believe that experimental investigation of the sub-Batchelor scale dynamics of the mixing process is an extremely interesting and urgent task.

\begin{figure}
 {\includegraphics[height=14cm]{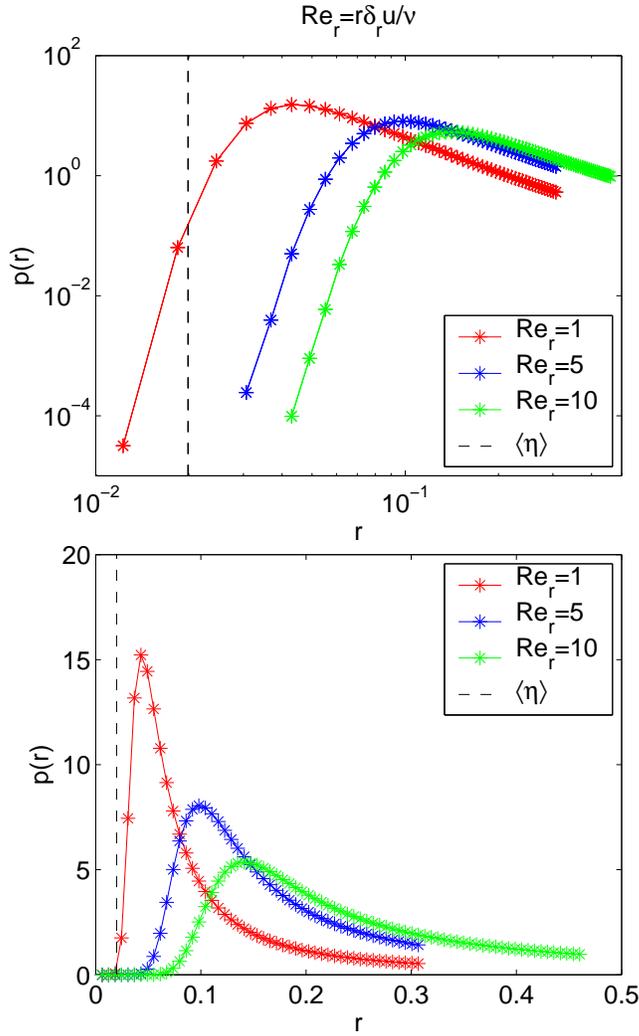}}
  \caption{ Logarithm of  not normalized probability densitiy   of  the disspation  scale  $P(r=\eta_{c})\equiv Q(\frac{\eta_{c}}{\eta_{0}},Re)$ for  $Re\approx 10^{3}$  and a few  different values of parameter $a$. 
The PDFs were numerically evaluated by Schumacher (2007) using  two different methods: 1.~Algorithm  based on the dissipation anomaly (eq. 2.1). (upper curve);  2~ From the isolevels of  dissipation rate ${\cal E}$.  (lower curve).  Broken lines denote Kolmogorov's scale. }
\end{figure}

\noindent The theory presented in this paper is based on the dissipation scale definitions  (2.1), (2.7),(2.8),  derived from the dissipation anomaly.  This algorithm was numerically  compared by Schumacher (2007) with the one based on the isosurfaces of the scalar dissipation rate.  The obtained PDFs $Q(\eta)$, though 
 qualitatively similar,   had quite substantial quantitative differences (see Fig. 3).   Since the relations based on dissipation anomaly (2.1) have been  derived directly from equations of motion, we believe they are  much better justified.

 \noindent  Interesting and stimulating discussions with N. Peters, U. Frisch, A. Kerstein,  J. Schumacher, E. Villermaux  and K.R. Sreenivasan are gratefully aknowledged.


\begin{thebibliography}{}
\bibitem{NonN} Batchelor,  G.K. (1959) , J. Fluid Mech. {\bf 5}, 113.
\bibitem{NonN} Bilger,  R.W., 2004,  {\em Some aspects of scalar dissipation},  Flow, Turbulence and Combustion {\bf 72}, 93-114.
\bibitem{NonN}.Buch K.A. \&  Dahm W.J., J. Fluid Mech. {\bf 364}, 1 (1998).
\bibitem{NonN} Celani,  A, Cencini, M, Vergassola, M, Villermaux, E., \& Vincenzi, D. 2005, {\em Shear effects on passive scalar spectra}, J.Fluid Mech. {\bf 523}, 99-108. 
\bibitem{NonN} Chertkov,  M., Falkovich, G. \&   Kolokolov, I. 1998,  Phys.Rev.Lett.{\bf 80},2121.
\noindent  Dimotakis,  P.E. 2005  {\em Turbulent Mixing},  Annu.Rev.Fluid Mech. {\bf 37}, 329-356 (2005);  {\em Some issues on turbulence and turbulent mixing}, CALCIT Report FM93-1. (March 1993). 
\bibitem{NonN}Duchon, J. \&   Robert,  R. (2000), Nonlinearity {\bf 13}, 249 
\bibitem{NonN} Eyink,  G.L. 2003, Nonlinearity {\bf 16}, 137 (2003).
\bibitem{NonN}  Gamba, A. \&   Kolokolov, I. 1999,  J.Stat.Phys. {\bf 94},   759.
\bibitem{NonN} Gotoh, T. \& Nakano, T. 2003, J. Stat. Phys.{\bf 113}, 855.
\bibitem{NonN} Kushnir, D.,   Schumacher,  \&  Brandt,  A. 2006,  {\em Geometry of intensive dissipation events in turbulence.},  Phys. Rev. Lett. {\bf 97}, 124502.
\bibitem{NonN} Landau, L.D. \& Lifshitz,  E.M. 1959, {\em Fluid Mechanics}, Pergamon Press, Oxford 1959.
\bibitem{NonN}  Monin, A.S. \& and Yaglom,  A.M., {\em Statistical Fluid
Mechanics, vol.\ 2}, MIT Press, Cambridge, MA .
\bibitem{NonN} Paladin, P. \&   Vulpiani,  A.  1987, Phys.Rep. {\bf 156}, 147 .
\bibitem{NonN}Peters,  N., 2000, {\em Turbulent Combustion}, (Cambridge University Press, Cambridge, England, 2000).
\bibitem{NonN} Polyakov, A.M. 1995,  {\em Turbulence without pressure.} Phys.\ Rev.\ E {\bf 52}, 6183.
\bibitem{NonN}  Schumacher, J.  \& Sreenivasan, K.R. (2003),  Phys. Rev. Lett., {\bf 91}, 174501.
\bibitem{NonN}  Schumacher, J. 2007, {\em Sub-Kolmogorov -Scale Fluctuations in Fluid Turbulence}, Phys.Rev.Lett.,  (submitted).
\bibitem{NonN} Schumacher, J, Sreenivasan, K.R. \& Yakhot, V. 2007, {\em Asymptotic exponents from low -Reynolds -number flows}, New. J. of Physics {\bf 9}. 
\bibitem{NonN} Sreenivasan, K.R. 2004, {\em Possible effects of small-scale intermittency in turbulent reacting flows}, Flow, Turbulence and Combustion {\bf 72}, 115-131.
\bibitem{NonN}  Tcheou, J.M.,  Brachet,  M.E.,  Belin,  F., Tabeling,  P., \&   Wiliaime, H. 1999, Physica D, {\bf 129}, 93-114.
\bibitem{NonN}  Yakhot , V. \&   Sreenivasan,  K.R.  2004, Physica A {\bf 343}, 147-155.
\bibitem{NonN} Yakhot, V. \&  Sreenivasan,  K.R. {\em Anomalous scaling of structure functions  and dynamic constraint on turbulence simulations},  J. Stat. Phys.{\bf 121} 823, ( 2005).
\bibitem{NonN} Yakhot,  V. 2006,  {\em Probability densities in strong turbulence}, Physica D {\bf 215}, 166, (2006).
\bibitem{NonN} Yakhot, V. 2003,  {\em Pressure-velocity correlations and anomalous exponents of structure functions in turbulence}, J. Fluid Mech.,  {\bf 495}, 135. 











 
 






 
\end{thebibliography}
\end{document}